\documentstyle[12pt,twoside]{article}

\renewcommand{\baselinestretch}{1.5}
\newcounter{numrom}
{\begin{list}{\Roman{numrom}.}
{\usecounter{numrom}}}{\end{list}}
\title{\bf Pion and Eta Photoproduction in Nuclei
        \protect\footnote[1]{Supported by BMFT and GSI Darmstadt} }
\author{A. Hombach, A. Engel\protect
\footnote[2]{Present address: Departement of Physics, Kyoto-University,
Kyoto, Japan.}
        , S. Teis and Ulrich Mosel \\
        Institut f\"{u}r Theoretische Physik, Universit\"{a}t Giessen \\
        D-35392 Giessen, Germany}

\begin{document}

\maketitle

\par

\begin{abstract}
We calculate the quasifree photoproduction of pions and eta mesons on nuclei
from $\mbox{}^{12}{\rm C}$ to $\mbox{}^{208}{\rm Pb}$.
Assuming that for the quasifree process the cross sections for
$\pi$/$\eta$ photoproduction on a nucleon in medium are
-- besides a lowering due to Pauli blocking -- identical to those on a free
nucleon, we use these as an input. The produced mesons or resonances then
propagate through the nucleus where they can undergo several scattering and
absorption processes. This final state interaction, which leads to a lowering
of the initial meson yield up to 70\%, we simulate in a coupled channel
Boltzmann-Uehling-Uhlenbeck transport model.
It allows us to study in detail the influence of Fermi motion, Pauli blocking
and of the additional decay channels for nucleon resonances in medium
on the photoproduction cross section.
The calculated total and differential cross sections are compared to
experimental data and the influence of the different medium effects on
the cross sections is discussed in detail.
\end{abstract}

\cleardoublepage
\end{document}


\documentstyle[12pt,twoside]{article}

\oddsidemargin 1.6cm
\evensidemargin 0.6cm
\textwidth14.0cm
\textheight21.0cm
\setlength{\parindent}{5mm}
\setlength{\parskip}{10pt}
\frenchspacing
\sloppy
\renewcommand{\baselinestretch}{1.5}
\addtolength{\topmargin}{-25pt}
\newcounter{numrom}
\newenvironment{enumrom}%
{\begin{list}{\Roman{numrom}.}
{\usecounter{numrom}}}{\end{list}}

\begin{document}

\def\figurename{Fig.}                        
\def\tablename{Tab.}                         
\makeatletter
\long\def\@makecaption#1#2{
 \vskip 10pt
 \setbox\@tempboxa\hbox{\bf#1: \small#2}   
 \leftskip=0mm\rightskip=0mm               
 \ifdim \wd\@tempboxa >\hsize
 \setbox3=\hbox{\bf#1: }                   
 \hangafter=1\hangindent=\wd3              
 \unhbox\@tempboxa\par \else \hbox
 to\hsize{\hfil\box\@tempboxa\hfil}
 \fi}
\makeatother

\renewcommand{\topfraction}{0.93}
\renewcommand{\bottomfraction}{0.93}
\renewcommand{\textfraction}{0.05}
\hyphenation{Aus-nutz-ung}
\hyphenation{Feyn-man}
\hyphenation{Dia-gram-me}
\hyphenation{Wech-sel-wir-kung}
\hyphenation{Form-faktor}
\hyphenation{Wirkungs-quer-schnitt}
\hyphenation{Selbst-ener-gie}
\hyphenation{Photo-produk-tion}
\hyphenation{Photo-produk-tions-wirkungs-quer-schnitt}
\hyphenation{Photo-produk-tions-wirkungs-quer-schnitte}
\hyphenation{Photo-produk-tions-wirkungs-quer-schnitten}
\hyphenation{Photo-produk-tions-wirkungs-quer-schnitts}
\hyphenation{Photo-produk-tions-wirkungs-quer-schnittes}
\hyphenation{Absorp-tions-wirkungs-quer-schnitts}
\hyphenation{Absorp-tions-wirkungs-quer-schnittes}
\hyphenation{Wel-len-funk-tion}
\hyphenation{Einteil-chen-wel-len-funk-tion}
\hyphenation{Teil-chen-wel-len-funk-tion}
\hyphenation{Test-teil-chen}
\newcommand{\enl}{\\[4pt]}
\newcommand{\be}{\begin{equation}}
\newcommand{\ee}{\end{equation}}
\newcommand{\bes}{\medskip\begin{equation}}
\newcommand{\ees}{\medskip\end{equation}}
\newcommand{\bea}{\begin{eqnarray}}
\newcommand{\eea}{\end{eqnarray}}
\newcommand{\beas}{\medskip\begin{eqnarray}}
\newcommand{\eeas}{\medskip\end{eqnarray}}
\newcommand{\aein}{\hspace*{.5cm}}

\pagenumbering{roman}
\cleardoublepage
\pagestyle{headings}
\pagenumbering{arabic}

\section{Introduction}

In the last years photoproduction of mesons on nucleons and
nuclei has attracted renewed interest due to new experimental facilities which
are expected to provide excellent data for both pion and eta production.
This allows the check of calculations and theories such as the LETs
for pion photoproduction on nucleons and the test of nuclear structure models
in the case of photoproduction on light nuclei.
In addition, photoproduction of mesons in nuclei is thought to provide a
test of resonance properties in the nuclear medium, because, first of all,
photons can excite nucleon resonances in matter of normal density
(in the center of heavy nuclei) and, second, because
the nucleus is left rather undisturbed and the background of signals from
other processes is small.
Consequently, many calculations have been performed in the past to
investigate photoproduction on nucleons, light nuclei and complex nuclei
\cite{car,car2,are86}; here we want to deal with the last ones.

The eta photoproduction on heavy nuclei has recently been studied by Carrasco
\cite{car}.
An effective Lagrangian model was used to calculate the primordial production
cross section, the final state interaction was
then simulated with a Monte Carlo (MC) code.
Assuming that eta photoproduction proceeds mainly via an excitation of a
$N(1535)$ resonance,
the $\eta NN^*$, $\pi NN^*$ and $\pi\pi NN^*$ coupling constants needed for
the Lagrangians have been fitted to the $\pi^- p \to \eta n$ data.
The decay ratios of the $N(1535)$ resonance were fixed to about 45\% for
$N^* \to N\eta$ and 55\% for \mbox{$N^*\to N\pi$},
$N\pi\pi$ according to the
particle data table \cite{par90}.
The $\gamma NN^*$ coupling is then determined by the experimental data for
the $\gamma p \to \eta p$ process.
{}From these Lagrangians a differential cross section for eta photoproduction
on a nucleon can be obtained. For nucleons inside nuclei medium corrections
are considered; the free $N(1535)$ width is replaced by a effective one due
to the new decay channels of the $N^*$.

The initial eta photoproduction on a nucleus is then given by a volume
integral over the nucleus whose density distribution is taken from
experiment, and an integral over the nucleon momenta obtained from a local
Fermi gas model.
A $\Theta$-function takes the Pauli blocking into account.
The absorption of the produced mesons is treated in a MC code.
{}From the $N(1535)$ selfenergy the
$\eta$ selfenergy is obtained, whose imaginary part gives the absorption
probability in each time step for an $\eta$ propagating though the nucleus.
According to the different partial widths of the exited $N(1535)$ resonance
it is then decided whether the $\eta$ scatters quasielastically or is
lost, that means, causes an \mbox{$N^*N\to NN$}
reaction or pion production from $N^*$ decay.

Pion photoproduction was investigated by Carrasco, Oset and Salcedo
\cite{car2}.
There mainly the same model as described above for eta photoproduction
was used. The calculation of the primary pion photoproduction cross section
is nevertheless more difficult than for the eta due to the strong nonresonant
contributions to the $\gamma N \to \pi N$ process.
For the pion propagation in the nucleus treated with a MC code two kind of
reactions are considered: Pion absorption including two- and three-body
absorption and quasielastic reactions including charge exchange.
Again the pion reaction probability is given by the pion selfenergy.
The angular dependence for both the $\gamma N \to N \pi$ and the
$\pi N \to \pi N$ reaction is calculated from the appropriate Feynman
diagrams
and also taken into account in the calculation of the initial angular
distribution
and the pion scattering in the MC code, respectively.

One of the most simple and earliest models for pion photoproduction was
presented by Arends et al. \cite{are86}.
Here the nucleus was treated as a Fermi gas and the free cross section
for $\pi^0$ photoproduction was used as an input. The final state interaction
is treated in a cascade calculation including $\pi^0 N$ scattering,
charge exchange and $\pi^0$ absorption by nucleon pairs.
The cross section for the last process is estimated from the
$\pi^+ d \to pp$ reaction and modified due to isospin and the density
ratio between deuterium and other nuclei.
In addition, the coherent photoproduction of $\pi^0$ was calculated,
which contributes, depending on the mass number, especially at lower
energies. The differential cross section for this process is calculated
by taking the spin independent part of the elementary
$\gamma N \to \pi^0 N$ cross section, and folding it with a nuclear matter
form factor and an absorption function depending on the mean free path
of a $\pi^0$. This cross section was then added incoherently to
the quasifree one.

Whereas all these models
treat the primary photoproduction step and its medium corrections
(e.g. corrections to resonance width) quite elaborately,
they all treat the final state interaction in a rather simplified manner
(e.g. neglecting $\eta N \to \pi N \to \eta N$ processes and the
propagation of the resonances).
We have, therefore, paid attention in particular to the last point,
and use the concept of quasi free particles in the moment of the
photoproduction process. Medium effects can correct the cross section,
but only after the resonances begin to propagate in the nucleus.

\section{The Model}

The BUU-model used in this work has been discussed in detail in many earlier
publications, e .g. \cite{cas90p,wol92}.
It is based on the Boltzmann-Uehling-Ulenbeck equation
\be
\frac{\partial f}{\partial t} + \frac{\vec p}{m} \frac{\partial f}
{\partial \vec r} + \vec \nabla U \frac{\partial f}{\partial \vec p}
= I_{\rm coll}
\ee
with $I_{\rm coll}$ representing a collision integral.
This equation describes the change of the phase space distribution  function
for the nucleons $f(\vec r, \vec p, t)$ due to collision processes and
propagation in a selfconsistent mean  field potential $U$.

For energies above the pion threshold, one has to include the inelastic
channels for the collisions, resonance creation and deexcitation or decay.
We are, therefore, using a coupled channel BUU method (CCBUU), that takes
nucleons, $\Delta$, $N(1440)$, $N(1535)$, $\pi$, $\eta$ explicitly into
account, including elastic $NN$ collisions and the inelastic processes
$NN \leftrightarrow NR$, $RR'\leftrightarrow NR''$, $R \leftrightarrow N \pi$,
$N(1535) \leftrightarrow N \eta$ ($R$ means resonance) \cite{wol92}.
The isospin degrees of freedom are considered, too.
The CCBUU equations are solved by the so called testparticle method in a
parallel ensemble algorithm \cite{ber88}. Due to the finite number of
testparticles a statistical error occurs; in our calculations we were
using 1000 testparticles per nucleon, leading to an uncertainty of the
results on the total cross section of about 6\%.

The treatment of the collision term is based on a geometrical
interpretation of the cross section.
Two particles can collide, if their relative distance is smaller then a
critical distance given by the cross section
\be
b_c = \sqrt{\frac{\sigma}{\pi}} \qquad.
\ee
For the calculation of the collision  distance and time the formalism of
Kodama et al. \cite{kod84} is used.
The cross sections are taken from experiment;
for $\Delta$ and $N(1440)$ creation the VerWest and Arndt parametrization
is used \cite{ver82},
the $N(1535)$ creation is parametrized according to Laget \cite{lag91}.
Resonance creation due to pion absorption is fitted to the $\pi N$ scattering
data and to the $\pi^- p \to \eta n$ reaction, respectively.
The $NR \to NN$ and $\eta N \to N(1535)$ cross sections are then calculated
via {\it extended detailed balance} \cite{wol92} according to
\be
\sigma_{NR\to NN} \,\,=\,\, S\,\,\frac{p_N^2}{p_R^2}\,\,\sigma_{NN\to NR}
\,\,\frac{1}{\int_{(m_N+m_\pi)^2}^{\sqrt{s} - m_N)^2} F(M^2)\,dM^2} \qquad,
\ee
where the factor $S$ has to be chosen appropriate to the situation
(identical particles in final state and spin average in initial state).
The integral in the denominator appears due to the fact that
the resonances are not given a fixed mass, but a mass distribution to
account for their finite lifetime in the case of $NN$ collisions;
in $\pi N$ and $\eta N$ reactions the resonance mass is determined by
energy-momentum conservation.

For the mean field potential a density dependent Skyrme parametrization
is used,
\be
U_N(\rho) = 0.75t_0\rho+\frac{7}{8}t_3\rho^\frac{5}{3}
\ee
with $t_0=-1177.8{\rm MeV/fm^3}$, $t_3=1845.5{\rm MeV/fm^3}$.
This leads to a binding energy for the nucleons of $-16$MeV at $\rho=\rho_0$
and a compressibility of the nuclear matter of 308.4MeV.
In addition a Yukawa potential with a range parameter of $2.175{\rm fm^{-1}}$
and strength $V_0 = -378{\rm MeV}$ is employed to obtain a better description
of the nuclear surface; it is
calculated in each time step by solving the appropriate Poisson equation.

A nucleus is initialized by first distributing the testparticles
according to a Woods-Saxon density distribution
\be
\rho(r)=\frac{\rho_0}{1+e^{(r-R)/a}} \qquad .
\ee
with the parameters $R=1.124A^{1/3}$fm and $a=0.0244\,A^{1/3} + 0.2864$fm.
This leads to a nucleus with minimum oscillations of density and rms.
In order to reduce the statistical fluctuations of the density due to
the finite number of testparticles, a testparticle does not only contribute
to the density at its position, but
weighted with a Gaussian function of 1 fm width also to the density at
neighbouring points.
The parameters for the potentials are adjusted such that the difference
between the experimental root mean square radius on the one side and
the two radii connected with the testparticle distribution and
the smeared density on the other is small.
Whereas the first one is smaller than the experimental one, the latter one
is slightly bigger. The differences are about 3\% (for $\mbox{}^{208}$Pb)
to 10\% (for $\mbox{}^{12}$C).
Density profiles for $\mbox{}^{12}$C and $\mbox{}^{208}$Pb
for both the pure testparticle distribution and
the smeared density in comparison to the experimental values are shown in
fig. \ref{1}.
While the measured density distribution of $\mbox{}^{208}$Pb is quite well
reproduced by both distribution functions, the curves for $\mbox{}^{12}$C
show a quite significant spread reflecting the inherent inaccuracies of
the semiclassical method when applied to light nuclei.

The momenta of the testparticles are uniformly distributed in
the local Fermi sphere whose radius is given by
\be
p_F(r) = \Bigl(\frac{3\pi^2}{2}\rho(r)\Bigr)^{\frac{1}{3}}
\ee
in a local Thomas Fermi approximation.

The initialization of the photon-nucleus reaction is illustrated in fig
\ref{2}.
Due to the small photon nucleon cross section no shadowing effects appear
in this energy range \cite{wei88}.
The reaction probability is therefore the same all over the nucleus,
at least for photon energies below 1 GeV.
So for the quasifree photoproduction process a testparticle is arbitrarily
chosen in each ensemble
and $\sqrt{s}$ and $\vec p_{cm}$ of the reaction are calculated;
here $\vec p_{cm}$ is the momentum of the photon-nucleon system in
the lab system where the nucleus is at rest.
Assuming that, e.g., for eta photoproduction a $N(1535)$ resonance is excited,
the resonance is given a mass equal to $\sqrt{s}$ and a momentum
$\vec p_{cm}$.
This resonance can now propagate through the nucleus and scatter, be absorbed
or decay according to the cross sections given in the coupled channel BUU
model.

For pion photoproduction the nucleus is initialized with a $\Delta$ only for
the $\gamma N \to \Delta \to N\pi$ part of the elementary cross section
(see next chapter). For the other channels the propagation of the resonance
state is skipped and the reaction is started directly with a nucleon-pion
pair with appropriate momenta taking Pauli blocking into account.
In case of a blocked reaction another testparticle of the ensemble
is chosen,
the number of nucleons checked unsuccessfully gives the Pauli blocking
factor of the reaction.

Pauli blocking in our model can be done in two different ways:
First, by checking numerically the occupation number of the phase space cell
of $(3{\rm fm} * 0.45{\rm fm}^{-1})^3$ for the nucleon final state; the
momentum direction for the nucleon is arbitrarily chosen in the
nucleon resonance rest frame.
In the second way, the number of the possible nucleon final states which
overlap with the local Fermi sphere and are thus blocked is determined
analytically.
We have verified that the total meson yield is changed by less than about
$\pm 10\%$ when going from one method to the other.

The number of mesons leaving the nucleus relative to the primary ones
determines the absorption probability.
Together with the Pauli blocking factor obtained when initializing the
nucleus directly with a pion-nucleon pair a factor is obtained with which
the elementary photoproduction cross sections have to be multiplied
in order to obtain the cross section on the nucleus.
\section{Pion Photoproduction}

For pion photoproduction on nucleons experimental data are available for
the channels $\gamma p \to n \pi^+$, $\gamma p \to p \pi^0$ as well as for
the $\gamma n \to p \pi^-$ reaction.
An analysis of the data shows that in the energy
range of interest from threshold up to approximately 1.4 GeV the two
dominant channels are the excitation of a $\Delta$ resonance and the
nucleon Born terms \cite{feu93,noz90}.
The $\Delta$ contributes to charged pion photoproduction even in the
resonance region only about 50\% whereas it is larger ($\sim$ 70\%)
for neutral pions (see e. g. fig. \ref{3}).
To take this fact into account the resonance part of the cross section was
parametrized separately, the difference to the total cross section taken as
Born part, thus neglecting any interference terms.
According to the ratios of these two cross sections, the nucleus was
initialized either with a $\Delta$ resonance or a pion-nucleon pair
(where the propagation of the intermediate state is neglected).
It turned out that this preparation of the initial state in two different
ways had only a small influence on the final results,
the difference between calculations performed with a pure $\Delta$- and
pure $\pi N$-pair - initialization was less than 10\%.
So also the neglect of the interference term between the resonance and
the Born contribution can play no significant role.

An additional result from the phase shift analysis \cite{feu93} and
the calculations of Garcilazo et al. \cite{gar93} is that the
$\gamma n \to n \pi^0$ cross section can be taken to be the same as the
$\gamma p \to p \pi^0$ one.
So no distinction was made between protons and neutrons for the
primary $\pi^0$ photoproduction.

The results of our calculations for neutral pion photoproduction are shown
in fig. \ref{4}.
For this study the light nucleus $\mbox{}^{12}{\rm C}$ was chosen
as well as the heavy $\mbox{}^{208}{\rm Pb}$.
In both cases the calculated total cross sections overestimate the data
at the peak energy of the $\Delta$ resonance and decrease faster at higher
and lower energies.
This behaviour is also seen in other calculations \cite{car2,are86,kmo84}.

In order to investigate why the calculations show a more pronounced
resonance structure than the data first the influence of the different
medium effects on the total cross section was analyzed, see fig. \ref{5}.
The strong resonance structure in the free $\pi^0$ photoproduction
cross section is smeared out for nuclei due to Fermi motion, but is still
clearly visible.
Pauli blocking then strongly suppresses the quasifree photoproduction at
lower energies thus shifting the effective production threshold up by more
than 50 MeV.
The absorption lowers the cross section by another factor of 3,
but does not change its shape significantly.
The deficiency of the calculations at photon energies above the
$\Delta$ resonance could thus be due to an overestimate of the
$\pi$-absorption at high energies
(see in this context the results of \cite{eng93}
which also give too much absorption in this energy range).
Alternatively, the suppression of a resonance structure in the data for
$\pi^0$ photoproduction on nuclei might be a hint for a strong broadening
of the resonance structure due to medium effects.
Experiments on photoabsorption \cite{bia94,from92} show an increase of the
effective $\Delta$ width; note, however, that the experimentally determined
width contains the effect of the Fermi motion.

In order to estimate the influence of multi-nucleon effects in pion
absorption discussed in the past (see e.g. \cite{sal87,eng93})
the $N\Delta \to NN$ cross section was modified in an ad-hoc way to
\be
\label{mod}
\sigma_{\Delta N}\,\,=\,\,(1+ \alpha\,\frac{\rho}{\rho_0})
\,\,\sigma_{\Delta N \to NN} \qquad,
\ee
thus simulating a three-body absorption term $(\sim\alpha)$;
four nucleon absorption should be negligible for
nucleon densities less than or equal to $\rho_0$.
Following the discussion in \cite{sal87}
$\alpha$ was chosen to be 1 which gives to the additional
three-nucleon absorption the same strength as to $\pi NN\to NN$.
The effects of this change on the total pion photoproduction cross section
were very small.
The structure was slightly broadened and the maximum value was only slightly
reduced.

This behaviour can be understood if one looks at the original creation point
of the final pions which is shown in fig. \ref{6}.
What can be seen is that most of the final pions are created in the
surface region of the nucleus where the density is already small.
Therefore the effect of three-nucleon processes is negligible
and the corrections to the free photoproduction cross sections
due to medium effects are small.
The pions from the center region of the nucleus, where the modification
(\ref{mod}) is most effective, are most likely absorbed.
This surface peaked creation of the final pions explains furthermore
the $A^{2/3}$ dependence of the total pion photoproduction cross section
\cite{are86} quite naturally (fig. \ref{7}).

The differential cross section for pion photoproduction on Pb is
shown in \mbox{fig. \ref{8}} for $\pi^+$ where experimental data exist
\cite{car2}.
The shape of the $\pi^0$ cross section is essentially the same, the maximum
value about 10\% higher.
Overall, the experimental angular distribution is reasonably well reproduced,
but the calculated distribution is somewhat too flat.

For completeness the anisotropy of both the $\gamma N \to N\pi $
and \mbox{$\pi N \to \pi N$} cross section was taken into account;
in the work of Engel et al. \cite{eng93} on pion-nucleon reactions
this had some influence for angle dependent cross sections.
In the case of pion photoproduction hardly any effect could be seen,
neither for the total nor for the differential cross section
unless the angular dependence for $\gamma N \to N\pi $ was made
artificially strong.
This can be understood because over 90\% of the pions have been rescattered
at least once before leaving the nucleus.
The angular dependence of the $\pi N \to \pi N$ cross  section is negligible
due to the high number of $\Delta N $ elastic collisions even in the
surface region.
This is only at first sight in contradiction to \cite{eng93};
for pion nucleus reactions it is the very first $\pi N$ scattering process
and its angular dependence which is decisive for the further development
of the reaction.

Processes which are so far not included in our model
are the coherent photoproduction and two pion creation.
Whereas the first one can only be treated in a completely different model,
the second one should be included in the future.

\section{Eta Photoproduction}

Eta photoproduction on nuclei was investigated in the photon energy range
from threshold at about 550 MeV to 900 MeV.
In this energy region the $N(1535)$ resonance is the only resonance with
an appreciable decay width into an eta meson of about 50 \%.
Furthermore, the dominant channel for $\eta$ photoproduction
is the excitation of a $N(1535)$ resonance \cite{feu93,tkb}.

To parametrize the $\eta$ photoproduction cross section,
the available data for $\gamma p \to \eta p$ have been fitted with a
Lorentz curve, see fig. \ref{9}.
This rather simple fit agrees well even with the new experimental
data from MAMI \cite{kru}.
For the $\gamma n \to \eta n$ process the same fit as for
the proton has been used with a reduced maximum value of 10$\mu$b .
This value was taken from a coupled channel analysis of the $\eta$
photoproduction on nucleons performed by Kamalov, Tiator and Bennhold
\cite{tkb}.
In their calculation including $N(1535)$ resonance, $\rho$ and $\omega$
exchange and nucleon Born terms, the assumption of a pseudoscalar coupling
of the $\eta$ to the nucleon best fitted the data for $\eta$ photoproduction on
protons.
Consequently, we used their result,
$\sigma_{\gamma n \to \eta n} \approx 1/2 \,\,\sigma_{\gamma p \to \eta p}$,
for the $\gamma n \to \eta n$ process for which no experimental data are
available, although there are indications that this
is in contradiction to data obtained from $\eta$ photoproduction
on the deuteron in MAMI experiments \cite{kru}.
Since the $N(1535)$ is a $S_{11}$ resonance, no angular dependence is to be
expected, neither for the elementary $\eta$ photoproduction process nor for
$\eta$ rescattering in the nucleus.

The results for $\eta$ photoproduction on nuclei are shown in fig. \ref{10}
for C and Pb.
The influence of Doppler broadening through Fermi motion, of Pauli blocking
and of reabsorption are shown in fig. \ref{11}.
A noticeable qualitative difference to the pion production case
(fig. \ref{5}) is the very steep rise of the elementary cross section at
about 710 MeV. This is due to the fact that at the energy threshold for
$\eta$ production the $N(1535)$ resonance is already well po\-pu\-la\-ted,
whereas for the $\pi^0$ case at threshold the $\Delta$ resonance carries
essentially zero strength and grows only slowly with increasing photon
energy. This different behaviour leads to the fact that the effect of the
Fermi motion of the target nucleus leads to a more significant lowering
of the effective production threshold for the $\eta$ than for the $\pi^0$.

The mass dependence for the total cross section, calculated
for a photon energy of 750 MeV and the four different nuclei
C, Ca, Sn and Pb, is found to be
$A^{2/3}$ (fig. \ref{12}), comparable to the pion photoproduction.
This increase of the total cross section with mass number is like
that of the nuclear surface; the latter determines the $A$-dependence
for a strongly absorbed particle.

In fig. \ref{13} we show the $\eta$ angular distribution for
the three energies \mbox{$E_\gamma = $ 710,} 750 and 790 MeV, for
$\mbox{}^{12}{\rm C}$.
All three of them peak at an angle of about $30^o$ and show a very steep
falloff
towards smaller angles and a slow drop towards larger angles.
This structure can be understood as follows (see also fig. \ref{14}).
A photon of 750 MeV energy impinging on a nucleon at rest would lead to a
very sharp angular distribution, peaked at $30^o$ in the lab.
Taking the Fermi motion into account leads to a broadening of the distribution
and, in particular, to a tail towards larger angles. This tail is due to
collisions of the incoming photon with nucleons whose momenta are
opposite to that of the photon. Pauli blocking and reabsorption,
also indicated in fig. \ref{14}, do not change this behaviour qualitatively,
and the low rescattering rate of the final $\eta$
(only about 15\% for Pb, 9\% for C)
makes this consideration to be valid also for $\eta$ photoproduction on nuclei.
At the peak angle of the differential cross section of $30^o$ also a
double differential cross section $d^2\sigma / d \Omega dp_\eta$
was calculated for $\mbox{}^{12}{\rm C}$ and a photon energy of 750 MeV.
It is shown in fig. \ref{15}.

An interesting question concerning $\eta$ photoproduction is the ratio
of a secondary $\eta$ production following an initial pion photoproduction.
The cross sections at least for charged pion production are about twice
as high as for the $\eta$ photoproduction; if there is enough energy
available for the pions they should be able to excite a $N(1535)$
resonance and thereby produce an $\eta$.
It has turned out that this primary pion photoproduction contributes
even in the case of Pb only about 12\% to the $\eta$ photoproduction
cross section, nearly independent of the photon energy.
This small contribution to the final $\eta$-yield is due to the kinematics
of the reaction;
already in the pion photoproduction step the nucleon takes too much kinetic
energy away, so the pions most likely excite a $N(1440)$ resonance
or a heavy $\Delta$ which cannot decay into an $\eta$ meson.
Pions from secondary decays are then even lower in energy.

\section{Conclusions}

We have investigated the photoproduction of $\pi$ and $\eta$ mesons on nuclei
and, in particular, have paid attention to the final state interaction of the
produced mesons.
We treat this final state interaction, which modifies the photoproduction
cross section significantly, in a coupled channel BUU model,
including nucleons, $\Delta$, $N(1440)$, $N(1535)$, $\pi$ and $\eta$.

Whereas Fermi motion leads to a wide broadening of the resonance structures
visible in the elementary cross sections, Pauli blocking and reabsorption
cause a strong suppression of the in-medium quasifree photoproduction
process at low energies.
The data for photo pion production are reasonably well reproduced, both in
their absolute magnitude and angular distribution.
However, the strong broadening of the $\Delta$ resonance peak in
the experimental data for pion photoproduction is not reproduced in our
calculations.
It is probably connected to problems concerning the photoabsorption
in nuclei, the broadening of the $\Delta$ resonance peak and the
missing absorption strength in the region of the $N(1440)$
resonance \cite{bia94}.
For $\eta$-production only preliminary data exist from the TAPS collaboration
\cite{kru}. It will be interesting to see how well our calculated total
cross section and angular distribution agree with them.
A problem in these calculations is the
uncertainty concerning the $\gamma n \to \eta n$ cross section,
since the calculated total cross sections on nuclei directly depend
on this input.

Though the primary photoproduction of mesons takes place in the whole
volume the experimentally observed mesons are most likely produced in the
surface region
of the nucleus. This leads to the observed mass dependence of the
total photoproduction cross section of $A^{2/3}$ for pions and
eta mesons.
Consequently, photoproduction experiments are
hardly usable for determining resonance properties in the nuclear medium;
the mesons created in the dense center region of the nucleus and
therefore most affected by a change of the resonance properties are most
likely absorbed.
The mesons created in the surface region can, however, only
show the influence of a low density zone.
This is shown also by the
negligible effect of the three-nucleon pion absorption
on the calculated pion photoproduction cross section.
Due to the strong surface dependence the
local Thomas Fermi approximation for the nucleon momenta used in our and
other calculations may be problematic.

Measurements of $\Delta$ properties in nuclei always
contain the broadening due to the Fermi motion of the nucleons.
So, additionally, we also checked the possibility to fix the initial
kinematical
situation of the $\gamma N -$reaction by detecting the recoil nucleons
together with the mesons.
It turned out that in the case of pion photoproduction, at $\gamma$ energies
of about 350 MeV, most likely two or three nucleons are kicked out;
in the case of higher photon energies even more.
This means, that the nucleon of the initial meson photoproduction process
collides quite often with the other nucleons around before
leaving the nucleus. The information about the momentum
direction and magnitude of the reaction partner of the photon
is destroyed by secondary scattering processes.

\section{Acknowledgement}

We gratefully acknowledge many helpful discussions with B. Krusche,
V. Metag and M. Rbig-Landau on the corresponding experiments presently
performed at MAMI.

We also thank the Institute for Nuclear Theory at the University of
Washington in Seattle for its hospitality and the Department of Energy
of the USA for partial support during the completion of this work.

\newpage

\renewcommand{\baselinestretch}{1.0}

\newpage

\begin{figure}
\vspace{2mm}
\caption{\label{1}
Density profiles for C and Pb. Shown are the testparticle distribution
(solid), the smeared density (dashed) and the experimental value obtained
from the charge distribution of the nuclei (dotted) \protect\cite{hof57}.
}
\end{figure}

\begin{figure}
\vspace{2mm}
\caption{\label{2}
Initialization of a photon-nucleus reaction.\hspace{5.2cm}
}
\end{figure}

\begin{figure}
\vspace{2mm}
\caption{\label{3}
$\gamma p \to n \pi^0$ process. Shown are the results of a phaseshiftanalysis
of \mbox{Th. Feuster} \protect\cite{feu93}: Contributions of the Born terms
(dashed), the $\Delta$ resonance (solid) and the higher resonances (dotted).
The data are taken from \protect\cite{lb88}.
}
\end{figure}

\begin{figure}
\vspace{2mm}
\caption{\label{4}
Total cross section of the $A(\gamma,\pi^0)A'$ reaction. The data are
taken from \protect\cite{are86}
}
\end{figure}

\begin{figure}
\vspace{2mm}
\caption{\label{5}
Photoproduction of $\pi^0$ on Pb, total cross section per nucleon.
Shown are the elementary cross section on a free nucleon (dotted), the cross
section with Fermi motion
of the nucleons (dash-dotted), the cross section with Fermi motion and
Pauli blocking (dashed) and the results of the full calculation,
including absorption of the produced mesons (solid).
}
\end{figure}

\begin{figure}
\vspace{2mm}
\caption{\label{6}
Original creation points of mesons that finally leave the nucleus.
Shown is the percentage of the final pions
at Pb with $E_\gamma = 350$ MeV in
comparison to the number of nucleons at this radial distance.
}
\end{figure}

\begin{figure}
\vspace{2mm}
\caption{\label{7}
Mass dependence of the total $\pi^0$ photoproduction cross section
at $E_\gamma = 350$ MeV. Calculated were the total cross sections for
C, Ca, Sn, Pb.
}
\end{figure}

\begin{figure}
\vspace{2mm}
\caption{\label{8}
Differential cross section for $\pi^+$ photoproduction on Pb with
$E_\gamma = 350$ MeV, $\vartheta$ is the lab angle.
The data are taken from \protect\cite{car2}.
}
\end{figure}

\begin{figure}
\vspace{2mm}
\caption{\label{9}
Parametrization of the $\gamma p \to \eta p$ process in comparison
to experimental data, taken from \protect\cite{lb88}.
}
\end{figure}

\begin{figure}
\vspace{2mm}
\caption{\label{10}
Total cross section for $\eta$ photoproduction on
C and Pb.\hspace{3.2cm}
}
\end{figure}

\begin{figure}
\vspace{2mm}
\caption{\label{11}
Photoproduction of $\eta$ on Pb, total cross section per nucleon.
Shown are the elementary cross section on a free nucleon (dotted),
the cross section with Fermi motion of the nucleons (dash-dotted),
the cross section with Fermi motion and
Pauli blocking (dashed) and the results of the full calculation,
including absorption of the produced mesons (solid).
}
\end{figure}

\begin{figure}
\vspace{2mm}
\caption{\label{12}
Mass dependence of the total photoproduction cross section for $\eta$
at $E_\gamma = 750$ MeV. Calculated were the cross sections for C, Ca, Sn, Pb.
}
\end{figure}

\begin{figure}
\vspace{2mm}
\caption{\label{13}
Differential cross section for $\eta$ photoproduction on C for different
photon energies: $E_\gamma = 710$ MeV (dotted), $E_\gamma = 750$ MeV (solid)
and $E_\gamma = 790$ MeV (dashed). $\vartheta$ is the lab angle.
}
\end{figure}

\begin{figure}
\vspace{2mm}
\caption{\label{14}
Differential cross section for $\eta$ photoproduction at $E_\gamma = 750$.
Shown is a scaled cross section for a nucleon at rest (dotted) in comparison
to different cross sections on Pb: Just with Fermi motion of the nucleons
(dashed-doubledotted), with Fermi motion and quasielastic rescattering of
the produced mesons (dashed-dotted), with Fermi motion, rescattering and
Pauli blocking (dashed) and including absorption of the produced $\eta$
(solid). $\vartheta$ is the lab angle.
}
\end{figure}

\begin{figure}
\vspace{2mm}
\caption{\label{15}
Differential cross section $\frac{d\sigma}{d\Omega d p_{\eta}} $.
This cross section was calculated at a scattering angle of $30^o$ in the
lab and a photon energy of 750 MeV for $\mbox{}^{12}$C.
$p_{\eta}$ is the lab momentum of the $\eta$ mesons.
}
\end{figure}

\cleardoublepage
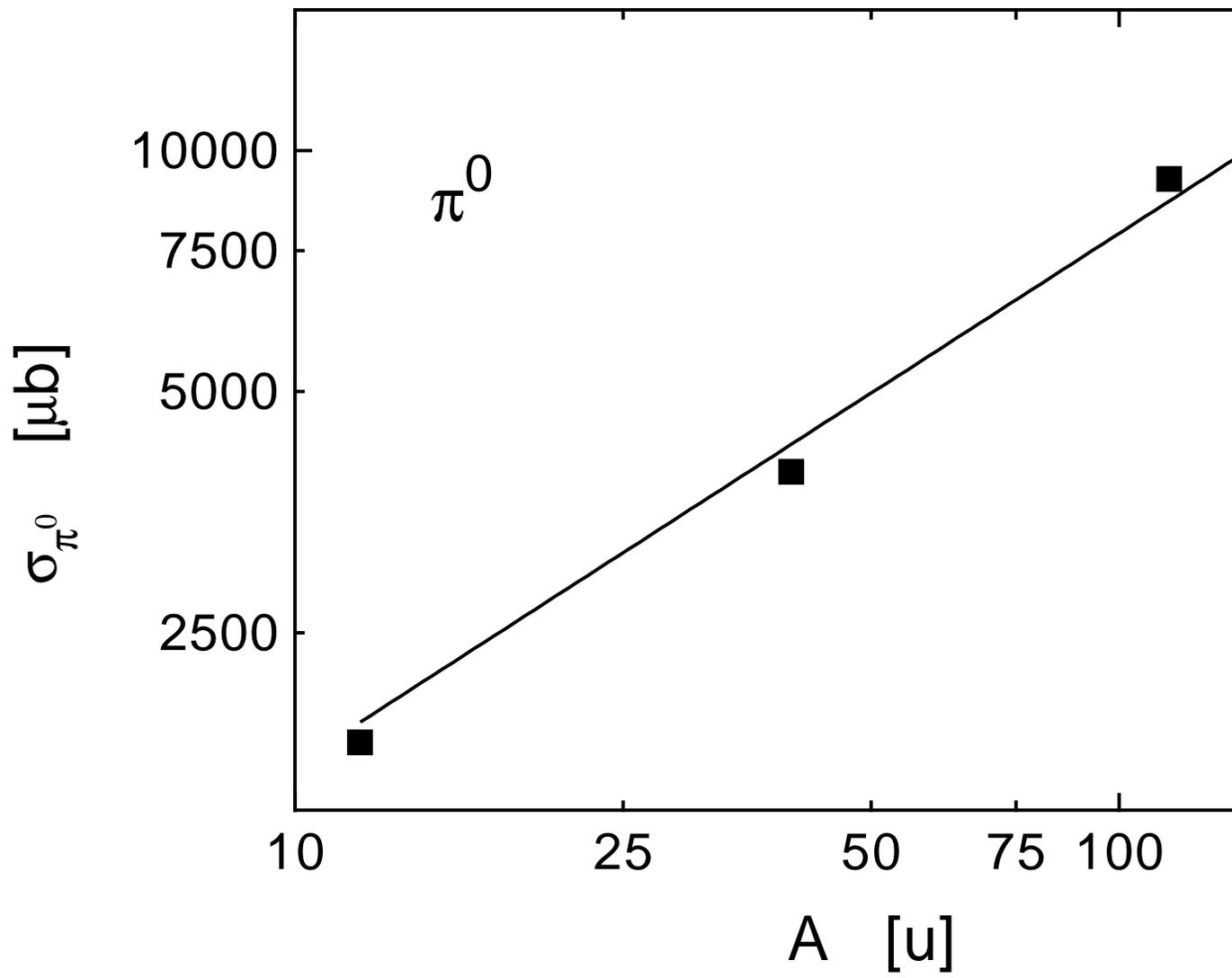
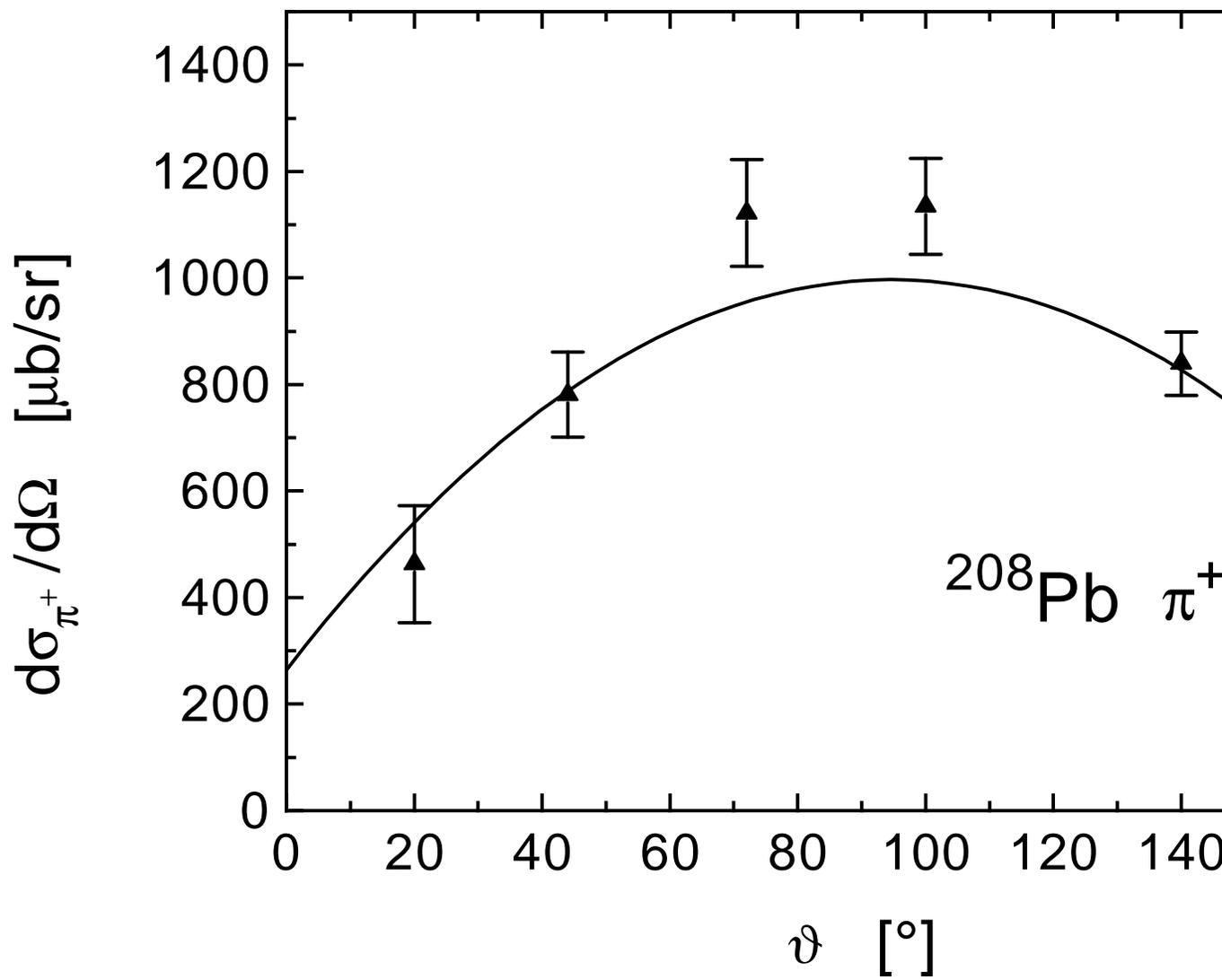
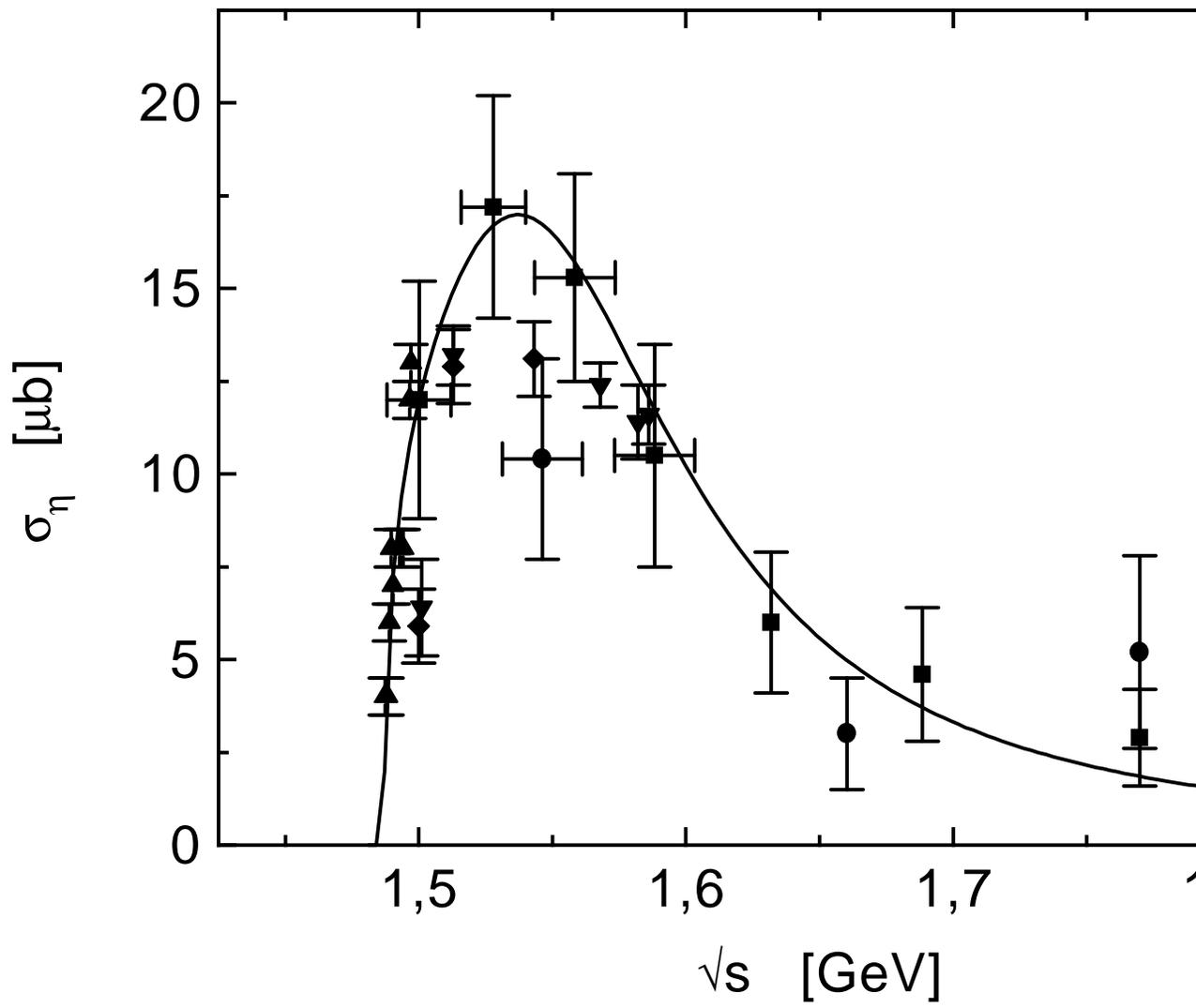
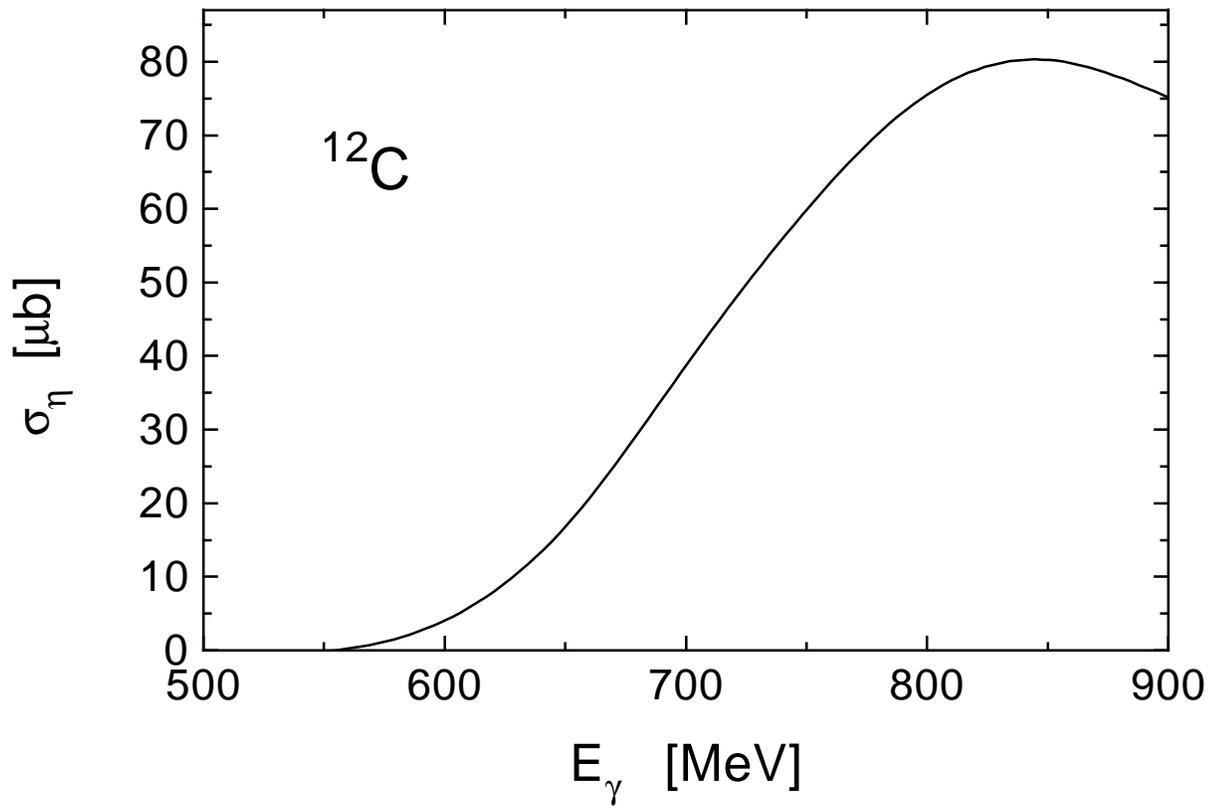
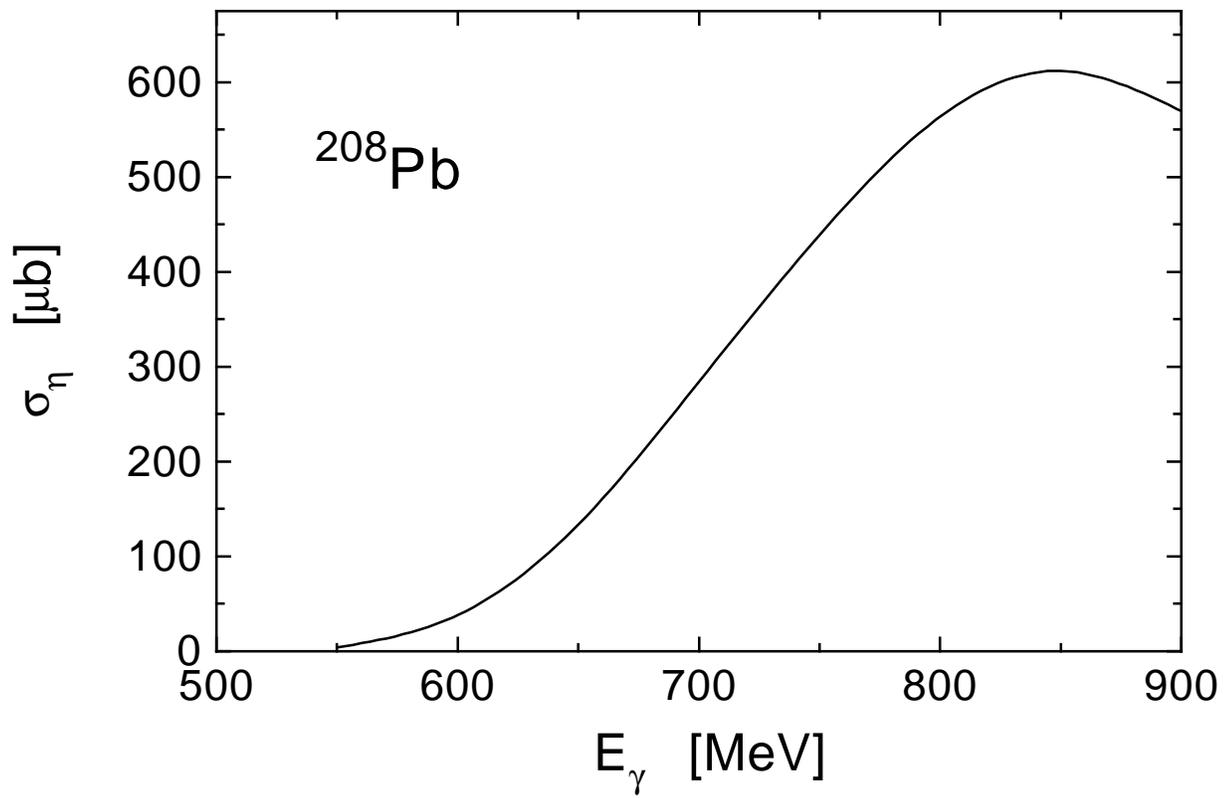
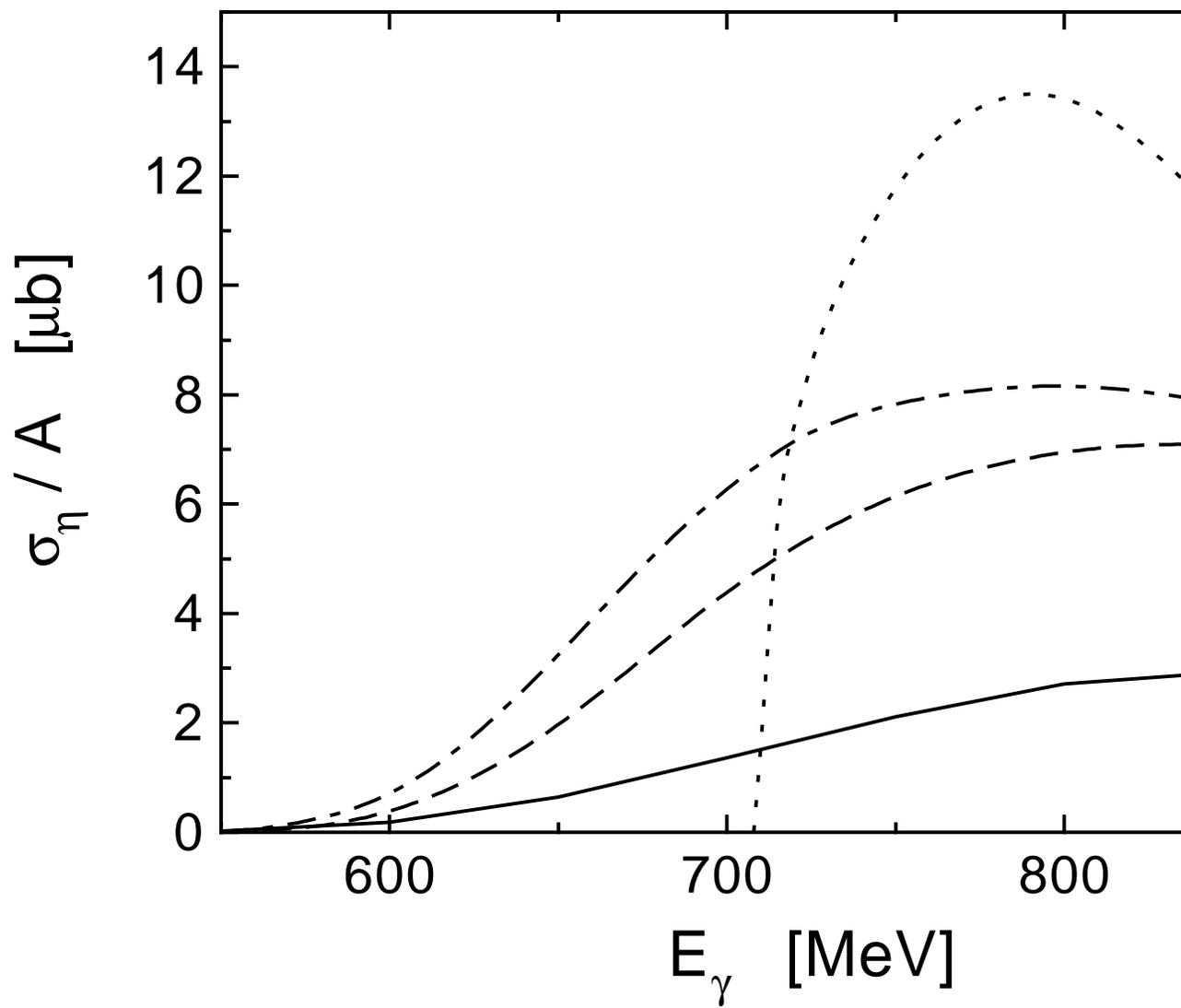
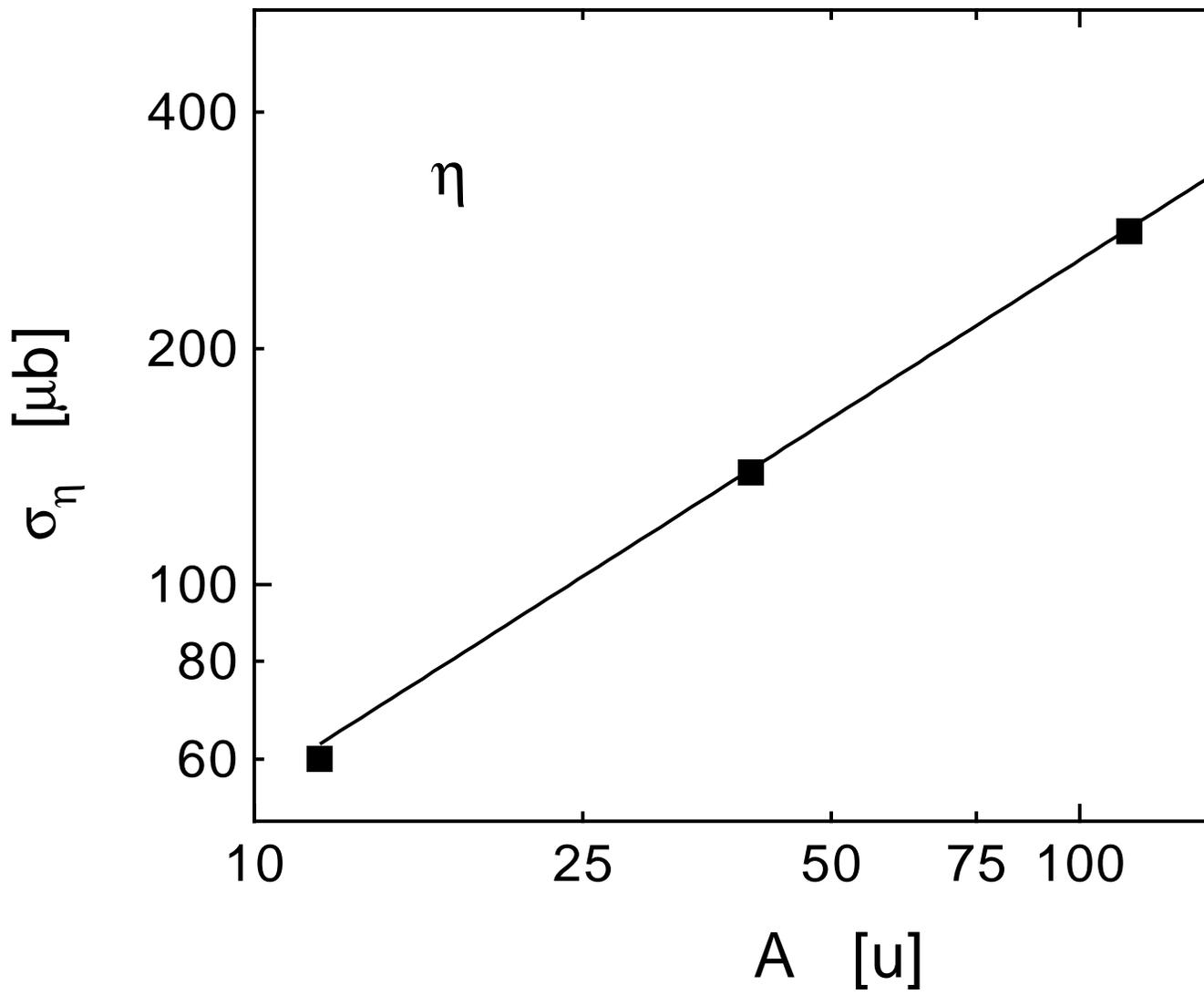
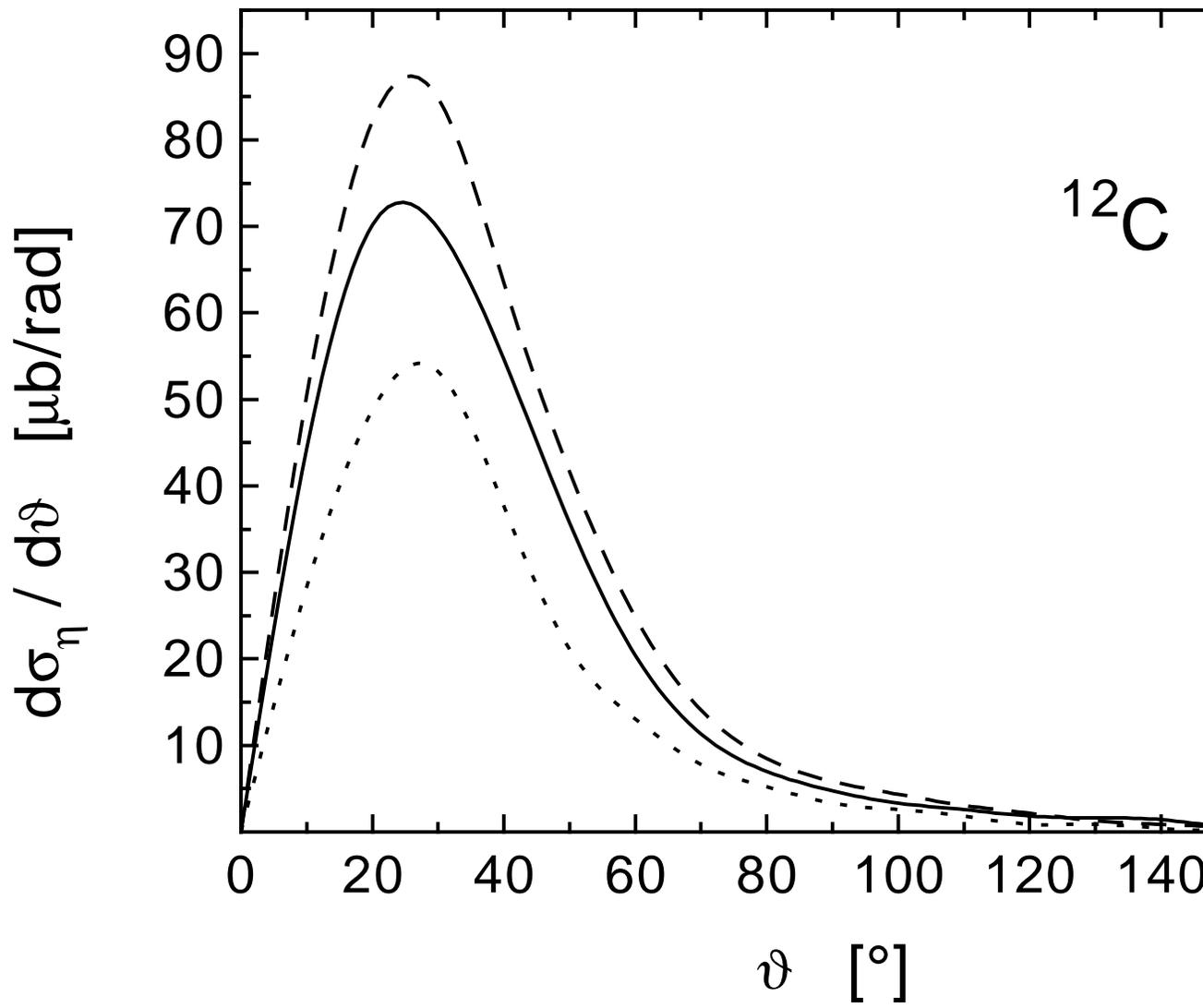
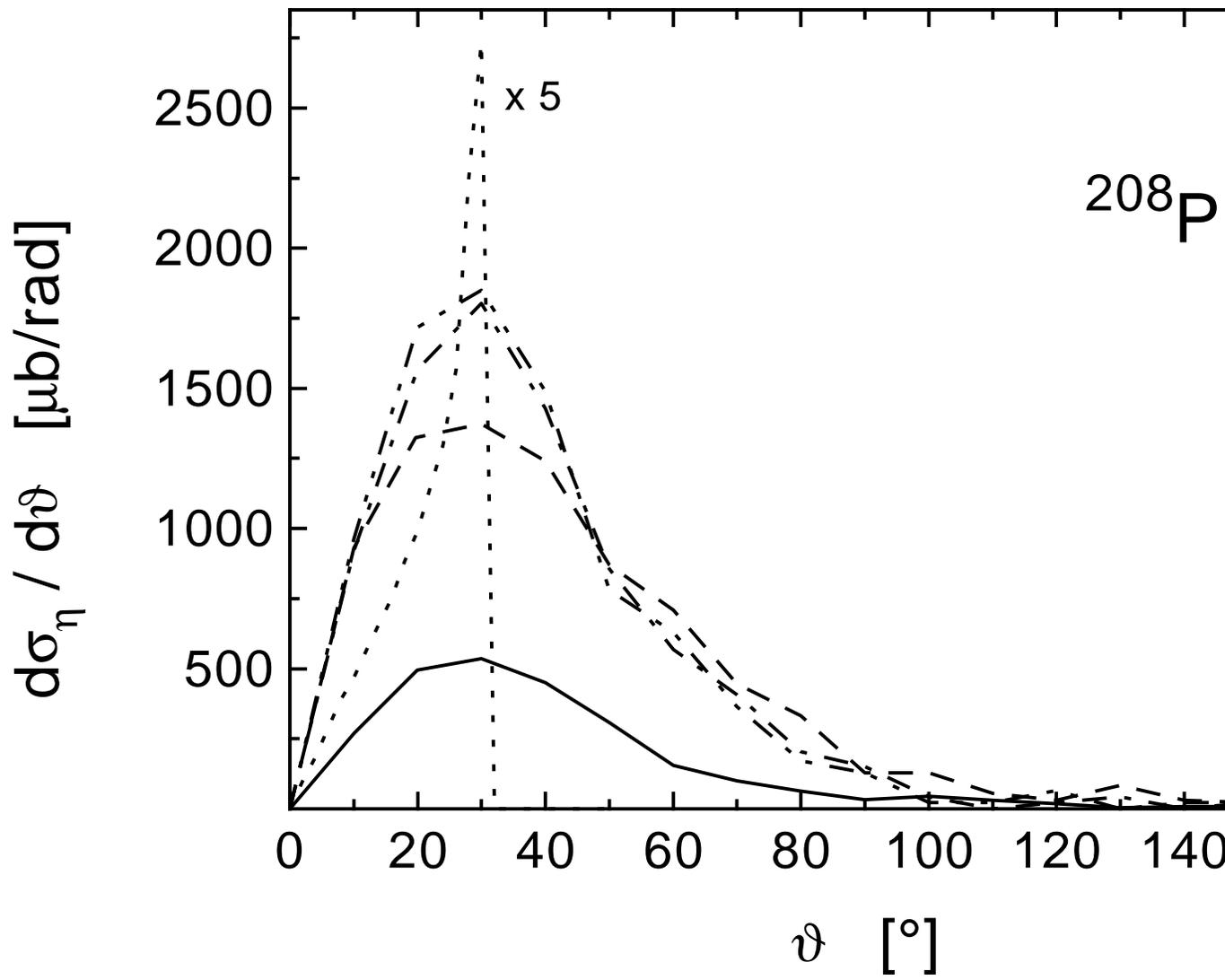
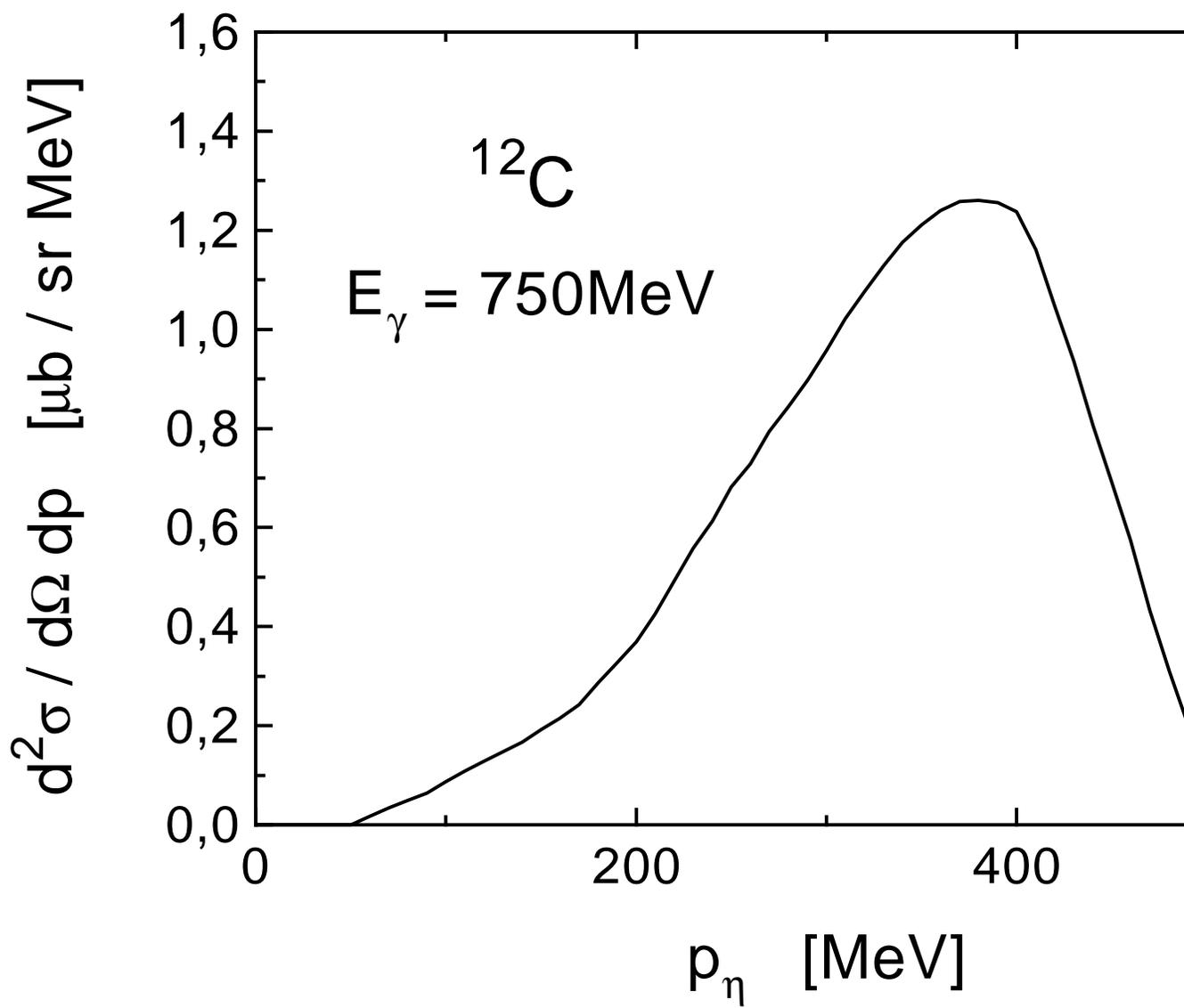
\end{document}